\documentclass[pra,superscriptaddress,nofootinbib,a4paper,twocolumn]{revtex4}
\usepackage{amsmath}
\usepackage{amsfonts}
\usepackage{graphicx}

\begin{document}

\title{Dialogue Concerning Two Views on Quantum Coherence: Factist
and Fictionist}
\author{Stephen D. Bartlett}
\affiliation{School of Physics, The University of Sydney, New South Wales 2006, Australia}
\author{Terry Rudolph}
\affiliation{Optics Section, Blackett Laboratory, Imperial College
London, London SW7 2BW, United Kingdom}
\affiliation{Institute for
Mathematical Sciences, Imperial College London, London SW7 2BW,
United Kingdom}
\author{Robert W. Spekkens}
\affiliation{Perimeter Institute for Theoretical Physics, 31
Caroline St. N, Waterloo, Ontario N2L 2Y5, Canada}
\date{31 October 2005}

\begin{abstract}
A controversy that has arisen many times over in disparate
contexts is whether quantum coherences between eigenstates of
certain quantities are fact or fiction. We present a pedagogical
introduction to the debate in the form of a hypothetical dialogue
between proponents from each of the two camps: a factist and a
fictionist. A resolution of the debate can be achieved, we argue,
by recognizing that quantum states do not only contain information
about the intrinsic properties of a system but about its extrinsic
properties as well, that is, about its relation to other systems
external to it. Specifically, the coherent quantum state of the
factist is the appropriate description of the relation of the
system to one reference frame, while the incoherent quantum state
of the fictionist is the appropriate description of the relation
of the system to another, uncorrelated, reference frame. The two
views, we conclude, are alternative but equally valid paradigms of
description.
\end{abstract}

\maketitle

\textit{This paper is dedicated to the memory of Asher Peres.
Asher had thought about and discussed with one of us many of the
issues we address here and had planned a paper of his own on the
subject. We will miss greatly the insight, clarity and
intellectual honesty he could bring to bear on the deepest
conceptual problems in physics.}

\section{Introduction}

We shall be considering a debate in quantum theory that has arisen
many times in many different contexts.  It is the debate over
whether it is possible to prepare states that are coherent
superpositions of eigenstates of certain observables.  One can
characterize the two camps by opposing slogans: ``coherence as
fact'' versus ``coherence as fiction''. Implicit in much of the
discussion is the assumption that quantum states directly describe
the intrinsic properties of a system and consequently that there
is a matter of fact about whether or not such coherences exist.

We shall suggest the following resolution to this debate: whether
or not it is appropriate to assume quantum coherences in the state
assignment for some system depends on the external reference frame
with respect to which that system is being described.
Specifically, it depends on whether this reference frame is
correlated with the system or not. We shall argue that the two
sorts of descriptions are both valid and consequently that the
presence or absence of coherences between eigenstates of certain
observables is not a matter of fact, but rather depends on one's
\emph{conventional} choice of reference frame. Central to our
argument is establishing the consistency of two descriptions: one
where the reference frame is treated \emph{internally}, in the
sense of receiving representation within the Hilbert space
formalism, and another where it is treated \emph{externally}, as a
classical system. Our position is by no means a new one; it has
many precursors in the literature, in particular in the work of
Aharonov and Susskind~\cite{Aha67}. We hope, however, that the
analysis presented herein will illuminate and add to what has come
before.

\section{The debate}

One context in which the debate over quantum coherence arises is
superconductivity, where there has been disagreement about whether
the Bardeen-Cooper-Schrieffer ground state, which involves a
coherent superposition of different charge eigenstates, is the
\emph{actual} state of a superconductor or whether the coherence
is merely a mathematical convenience~\cite{And86,Haa62,Ker74}. The
same argument arises in the context of Bose-Einstein condensation
regarding coherent superpositions of different atom number
eigenstates~\cite{Jav96,Hos96,Jav97,Cas97}. In both cases,
standard practice in the condensed matter community is to assign
an order parameter to the condensate, which is typically defined
as the expectation of the quantum field operator. Thus, if the
quantum state has no coherence between different eigenstates of
the number of Cooper pairs or atoms, the order parameter is zero.
However, the usefulness of the concept of a non-zero order
parameter suggests that there is something wrong with this
approach. The prediction~\cite{Jav96} and subsequent
observation~\cite{And97} of the interference of independent atomic
condensates has also fueled the debate on phase coherence in these
systems.

The debate has also arisen in the context of quantum optics, where
the issue is the existence of coherence between eigenstates of
different photon number.  This is the forum in which the debate
has seen the most recent activity (although an early version of it
can be found in Refs.~\cite{Har94,GHZ95,Har95}). It is also
perhaps the context in which the debate has been the most
sophisticated, due to the advanced techniques for describing and
implementing generalized measurements and state preparations now
commonly employed by the quantum optics community.  Also, in as
much as the physical descriptions in quantum optics can be
directly derived from a fundamental theory (Quantum
Electrodynamics) without recourse to effective theories, as in the
Bose-Einstein condensation and superconductivity examples, one
might have expected that such controversies can be rigorously
settled one way or the other.  We will focus on the optical
context here for concreteness.

Recent interest in the optical version of the debate begins with the
1997 paper by Klaus M\o lmer entitled ``Optical coherence: a
convenient fiction''~\cite{Mol97}. The standard assumption in the
quantum optics community is that a laser operating above threshold
emits an electromagnetic (EM) field for which the quantum state is
\begin{equation}
\left\vert \alpha \right\rangle =\sum_{n=0}^{\infty }\frac{e^{-|\alpha
|^{2}/2}\alpha ^{n}}{\sqrt{n!}}\left\vert n\right\rangle
\label{coherentstate}
\end{equation}
where $\alpha $ is complex. This is known as a \emph{Glauber state}
or \emph{coherent state}. It is a coherent superposition of photon
number eigenstates with a phase that varies linearly with number,
and number-state populations that obey a Poissonian distribution. In
other words, if $\alpha =\sqrt{\bar{n}}e^{i\phi }$, then the
relative phase between $\left\vert n+k\right\rangle $ and
$\left\vert n\right\rangle $ is $e^{ik\phi }$, and the probability
of $n$ photons is
\begin{equation}
  p_{n}=\frac{e^{-\bar{n}}\bar{n}^{n}}{n!}\,.
  \label{Poisson}
\end{equation}
In his paper, M\o lmer tries to cast doubt on this assumption
about the state of a laser by considering how a laser field is
produced.

His argument relies on the following assumptions: (i) the atoms of
the gain medium are treated quantum mechanically, (ii) these atoms
are initially described by an incoherent mixture of energy
eigenstates, and (iii) energy is conserved in the interaction
between the atoms and the optical field. With these assumptions,
the interaction between an atom in the gain medium and the
electromagnetic field is such that, if the atom is excited and the
field is initially described by an $n$ photon eigenstate, then the
atom+field evolves over a time $t$ to a coherent superposition of
what one started with and a state wherein the atom has de-excited
and the field has acquired an additional photon,
\begin{equation}
  \left\vert e\right\rangle \left\vert n\right\rangle \rightarrow
  a(t)\left\vert e\right\rangle \left\vert n\right\rangle +b(t)\left\vert
  g\right\rangle \left\vert n+1\right\rangle \,,
\end{equation}
where $a(t)$, $b(t)$ are complex amplitudes. Note that this state is pure
and entangled. If one is interested only in the reduced density operator of
the field, obtained by taking the trace over the atom, one finds that the
state of the field is an incoherent mixture of $n$ and $n+1$ photons,
\begin{equation}
\rho =|a(t)|^{2}\left\vert n\right\rangle \left\langle n\right\vert
+|b(t)|^{2}\left\vert n+1\right\rangle \left\langle n+1\right\vert .
\end{equation}

The gain medium of the laser as a whole is simply an incoherent
sum of different numbers of excitations, each term of which
evolves to an entangled state between the gain medium and the
field. A careful analysis~\cite{Mol97} shows that the reduced
density operator of the field is found to be of the form:
\begin{equation}
  \rho =\sum_{n=0}^{\infty }p_{n}\left\vert n\right\rangle \left\langle
  n\right\vert \,,
  \label{incoherentstate}
\end{equation}
with $p_{n}$ the Poissonian distribution of Eq.~(\ref{Poisson}).
Thus, although the populations of the number states are what we
expected (the same as for the coherent state), there are no
coherences, and thus no phase relations, between these. Thus,
surprisingly, M\o lmer's account of the inner workings of the laser
seems to imply that the field emitted by a laser operating above
threshold is not the coherent state of Eq.~(\ref{coherentstate}) as
is usually assumed, but rather the incoherent state of
Eq.~(\ref{incoherentstate}).

M\o lmer concludes that the coherence was just a fiction. Maybe it is
convenient to assume, maybe one doesn't make mistakes by assuming it, but it
isn't \emph{really} there.

\section{A possible dialogue}

Subsequent to this, there was a flurry of activity on the
subject~\cite{Gea98,Mol98,Rud01a,vEF02a,Rud01b,vEF02b,Nem02,Spe03,San03,Wis03,Wis04,Fuj03,Smo04}.
In addition to what has appeared in the literature, there have been
a great number of debates on this issue at various conferences and
among workers in the field, which supplement the arguments found in
the literature. We have ourselves benefitted a great deal from early
discussions with John Sipe, and ongoing discussions with Barry
Sanders and Howard Wiseman on this subject. There is much to learn
from the details of these debates. Note that we do not attempt to
provide a historically accurate account of the relevant literature
or of who believed what at various stages of the debate. Rather, we
shall try to simply give a flavour of the argument and the central
issues. We therefore present the debate in the form of a
hypothetical dialogue between purists from the two camps. \ This
dialogue is representative of many of the arguments and
counterarguments that have been provided.\footnote{ The arguments
for both sides of the debate are sufficiently compelling that it is
easy to find oneself (and indeed we have found ourselves) defending
different positions at different times. }

We shall call the proponent of the idea that coherence is fact
``the factist'' and the proponent of the idea that coherence is
fiction ``the fictionist''. We join the story at the stage where
the fictionist is just finishing the argument we described above.

\textbf{Fictionist:} [...] And so you see, if you do a proper quantum
analysis of the manner in which laser light is produced, you find that the
reduced density operator of the field is an incoherent sum of number states,
not a coherent superposition. Optical coherence is a fiction!

\textbf{Factist}: It's a cute argument, and I admit that I had to
give it some thought before I saw what was wrong with it, but I've
figured it out. The key is that the reduced density operator can
be written as an incoherent sum of number states, but it can also
be written as an incoherent sum of coherent states. That is,
expressing $\alpha =|\alpha |e^{i\phi }$ in polar coordinates,
with $\phi $ the phase of the coherent state $|\alpha \rangle =|
{|}\alpha |e^{i\phi }\rangle $, then we can rewrite
Eq.~(\ref{incoherentstate}) as
\begin{equation}
\rho =\sum_{n=0}^{\infty }p_{n}\left\vert n\right\rangle \left\langle
n\right\vert =\int_{0}^{2\pi }\frac{d\phi }{2\pi }\left\vert |\alpha
|e^{i\phi }\right\rangle \left\langle |\alpha |e^{i\phi }\right\vert \,.\end{equation}
As a result, we can interpret the situation as follows: the field
is \emph{actually} in some particular coherent state $|\alpha
\rangle $, we just don't know which, because we don't know \emph{a
priori} what is the phase of the laser. As a result of this
ignorance, we have to represent our knowledge by an incoherent sum
of coherent states equally weighted over all phases. So the old
way of looking at this was right all along. There really is
coherence there.

\textbf{Fictionist}: Look, I know all about this multiplicity of
convex decompositions of a mixed state, sometimes called the
``ambiguity of mixtures'' \cite{Sch36,HJW93}, but you're wrong to
assign special significance to one such decomposition because one
cannot adopt an ignorance interpretation of an improper mixed
state.\footnote{ An improper mixture is one that arises as the
reduced density operator of a pure entangled state, while a proper
mixture is one that arises as an incoherent sum of pure
states~{\cite{dEspagnat}.}} Here's the problem. If you tell me
that \emph{really} the quantum state of the field is $|\alpha
\rangle $ for some $\alpha $, then you're telling me that
\emph{really} the reduced density operator for the field is
$|\alpha \rangle \langle \alpha |$ and the only quantum state of
the atoms+field that has this as a reduced density operator is a
product state of the form $\left\vert \chi \right\rangle
\left\vert \alpha \right\rangle$ for some atomic state $\left\vert
\chi \right\rangle$. But then you are saying that \emph{really}
the atoms+field system is in a product state, and this contradicts
the assumption we started with, that the atoms+field system is in
an entangled state.

\textbf{Factist}: I suppose I hadn't thought that through
carefully enough. But now that I have, I realize what the correct
response is. One of your assumptions was that the gain medium of
the laser was in an incoherent mixture of energy eigenstates, but
you're wrong. It's actually in a coherent superposition of energy
eigenstates. Roughly speaking, the lasing phase transition occurs
because the atoms start oscillating in phase with each other due
to a symmetry breaking which occurs when stimulated emission
(which preserves phase) dominates spontaneous emission - and we
must therefore describe the atomic state $|\phi \rangle$ as one
depending on the common phase $\phi$ of their oscillation. Simple
mean field theory descriptions of the symmetry breaking
accompanying this transition \cite{DeG70,Gra74} show that the
state of the atoms is one which involves a non-zero expectation
value of the atomic dipole moment operator, which in turn implies
that their state involves a coherent superposition of energy
eigenstates. The standard atom-field interaction serves to
transfer this coherence to the emitted field - that is why there
is a nonvanishing expectation of the annihilation operator for the
field.

\textbf{Fictionist}: \emph{Even if} the gain medium had a well-defined
phase, you don't know what it is, so you have to describe it by the state
that is a mixture over all phases,
\begin{align}
  \rho _{\text{atoms }} &=\int_{0}^{2\pi }
  \frac{d\phi}{2\pi}\left\vert \phi
  \right\rangle \left\langle \phi \right\vert  \\
  &=\sum_{n}w_{n}\left\vert E_{n}\right\rangle \left\langle
  E_{n}\right\vert \,,
\end{align}
where $\left\vert E_{n}\right\rangle$ are the energy eigenstates
of the gain medium and $w_{n}$ is some probability distribution.
It is $\rho_{\text{atoms}}$ that you should use in the
calculation, and because this state has a vanishing dipole moment,
you can't develop any coherence in the field, as I showed before.

\textbf{Factist}: The density operator $\rho_{\text{atoms}}$ might
be fine for calculations, but what's \emph{actually going on} is
that there is some pure state $\left\vert \phi \right\rangle $
that describes the atoms.

\textbf{Fictionist}: [\textit{groan}] Haven't we been over this
before? For you to interpret $\rho _{\text{atoms }}$ as a mixture
of the states $\left\vert \phi \right\rangle $ rather than a
mixture of the states $\left\vert E_{n}\right\rangle $ is to favor
one convex decomposition over another, which is a fallacy!

\strut \textbf{Factist}: Not this time! It's only a fallacy if the
system in question is really entangled with something else, which
is not the case for the atoms of the gain medium.

\textbf{Fictionist}: Well, \emph{really}, the gain medium is
prepared by some pumping mechanism, and we need to treat the
electrons of this mechanism quantum mechanically. Because they
start out in a proper mixture of energy eigenstates with no dipole
moment and because the interaction with the atoms of the gain
medium is energy-conserving to good approximation, we find that in
the end the electrons and the atoms are in a proper mixture of
entangled states. Each such entangled state is a coherent
superposition of different ways of distributing the energy between
the pumping mechanism and the gain medium and the reduced density
operator on the atoms for each is an \emph{improper} mixture. So
to claim that one convex decomposition of $\rho_{\text{atoms}}$ is
preferred is indeed a fallacy.

\textbf{Factist}: I question your assumption about the initial state
of the pumping mechanism, but I'm going to drop it because it seems
to me that the question of whether or not there exists coherence
should ultimately be settled by experiment, and unfortunately for
you, the experiments show that the fields emitted by lasers
\emph{are} in coherent states.

\begin{figure}
\begin{center}
\includegraphics[width=2.5in]{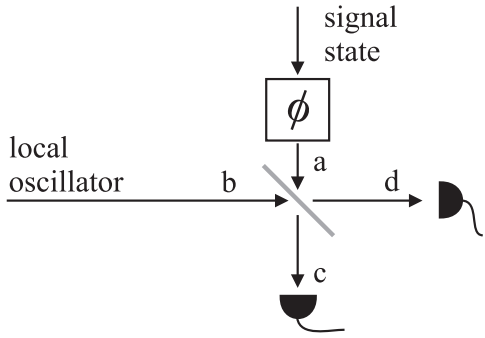}
\end{center}
\caption{Factist's schematic for homodyne detection of a signal
state using a local oscillator, a beamsplitter, and photodetectors.
A variable phase shift $\phi$ can be applied to the signal state.}
\label{homodyne1}
\end{figure}

The experiment I'm thinking of is a simple balanced homodyne
detection [\textit{shown in Fig.~\ref{homodyne1}}]. One mixes the
signal (mode $a$) with a local oscillator (mode $b$) at a 50/50
beam splitter and detects the difference in the intensity at the
two output ports (modes$\ c$ and $d$). \strut The signal is given
a variable phase shift of $\phi $ prior to the beam splitter, and
the difference in intensity is measured as a function of this
phase shift.

Suppose that the path lengths are arranged such that for $\phi =0$
one finds the mean intensity at the two output ports to be equal. If
the signal mode is initially described by the density operator
$\rho$, then after a phase shift of $\phi$, it is described by
$e^{i\phi N}\rho e^{-i\phi N}$ where $N$ is the number operator. Let
$a$ denote the annihilation operator for the signal mode, let $\beta
$ denote the classical electromagnetic field for the local
oscillator, and let $c$ and $d$ denote the annihilation operations
for the two output modes. By the action of the beam splitter,
\begin{equation}
  c=\frac{1}{\sqrt{2}}\left( a-\beta \right) \qquad
  d=\frac{1}{\sqrt{2}}\left( a+\beta \right) .
\end{equation}
The Hermitian operator associated with the difference in intensity
at the output ports is~\cite{Leo97}
\begin{equation}
d^{\dag }d-c^{\dag }c=  (\beta^* a+\beta a^{\dag })
\end{equation}
Thus, for a quantum state $\rho$ that suffers a phase shift of
$\phi$ we expect an intensity difference of
\begin{align}
  I_{d}-I_{c}  &= \text{Tr}\left( e^{i\phi N}\rho e^{-i\phi N}
  (\beta^* a+\beta a^{\dag })\right)  \\
  &= \text{Tr}\left( \rho  (\beta^* e^{-i\phi }a+ \beta e^{i\phi }a^{\dag
  })\right) \,.
\end{align}
Interference, that is, variation of $I_{c}-I_{d}$ with $\phi ,$ can
only arise if $\mathrm{Tr}(\rho a)$ is non-zero, that is, if $\rho$
has coherence between different number eigenstates.

Thus, seeing interference in a homodyne detection measurement
demonstrates the presence of coherence. The experiment has been done
for a coherent state signal, $\rho =|\alpha\rangle\langle\alpha|$,
and the interference is observed. So \emph{experiment} shows that
coherences exist.

\begin{figure}
\begin{center}
\includegraphics[width=3.25in]{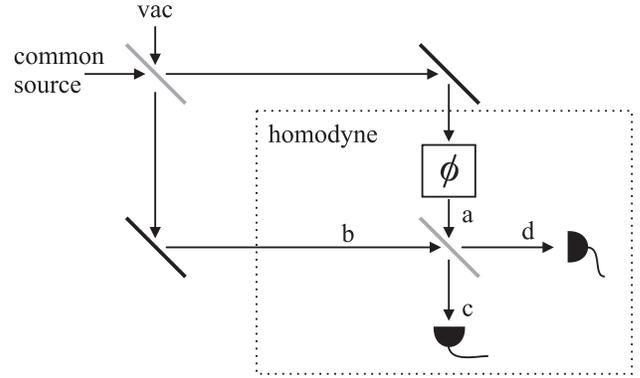}
\end{center}
\caption{Fictionist's schematic for homodyne detection wherein the
local oscillator arises from the same common source as the
signal.} \label{homodyne2}
\end{figure}

\textbf{Fictionist}: \ [\textit{Shaking her head}] You're always
forgetting to think a bit about where these systems (such as your
``local oscillator'') come from! The experiment actually looks like
Fig. \ref{homodyne2}.

The local oscillator and the signal are not independent -- they come
from a common source! So really what happens is that the source is
beat against the vacuum at an unbalanced beam splitter with
transmission probability $T$. If the source is a Fock state
$|n\rangle $, then after the first of the two beam splitters, the
state of modes $a$ and $b$ is~\cite{Wal94}
\begin{equation}
  \label{psi_n_phi}
  \left\vert \psi _{n}\right\rangle
  =\sum_{m}c_{m}^{(n)}\left\vert m\right\rangle \left\vert
  n-m\right\rangle \,,
\end{equation}
where
\begin{equation}
  c_{m}^{(n)}=2^{-n/2}\sqrt{\binom{n}{m}}T^{m/2}(1-T)^{(n-m)/2}.
\end{equation}
The phase shift by $\phi $ causes the quantum state to evolve to
\begin{equation}
  \left\vert \psi _{n,\phi }\right\rangle
  =\sum_{m}c_{m}^{(n)}e^{-i\phi m}\left\vert m\right\rangle
  \left\vert n-m\right\rangle ,
\end{equation}
But note that the reduced density operator for the signal is still of the
form
\begin{equation}
\rho =\text{Tr}_{b}\left\vert \psi _{n,\phi }\right\rangle
\left\langle \psi _{n,\phi }\right\vert
=\sum_{m}|c_{m}^{(n)}|^{2}\left\vert m\right\rangle \left\langle
m\right\vert \, \\
\end{equation}
which is devoid of coherence.

It is of course more realistic to assume that the source puts out a
Poissonian mixture of Fock states, but then after the beam splitter and
phase shifter, the state is
\begin{equation}
\label{rho_ab}
\rho _{ab}=\sum_{n}p_{n}\left\vert \psi _{n,\phi
}\right\rangle \left\langle \psi _{n,\phi }\right\vert \,.
\end{equation}
The reduced density operator for the signal mode is
\begin{align}
  \rho =\text{Tr}_{b}(\rho
  _{ab})&=\sum_{m}(\sum_{n}p_{n}|c_{m}^{(n)}|^{2})\left\vert
  m\right\rangle \left\langle m\right\vert \, \\
  &=\sum_{m} p_{m}\left\vert m\right\rangle \left\langle
  m\right\vert \,,
\end{align}
where $p_m$ is a Poissonian distribution (the proof of this is
straightforward using the techniques of Ref.~\cite{San03}).  Thus,
the state of the signal is incoherent.

Despite this incoherence, one still predicts interference.
Denoting the annihilation operator associated with the local
oscillator by $b$, we have
\begin{equation}
  c=\frac{1}{\sqrt{2}}\left( a-b \right), \qquad
  d=\frac{1}{\sqrt{2}}\left( a+b \right)\,.
\end{equation}
The Hermitian operator corresponding to the relative number of
photons found at the two detectors is therefore
\begin{equation}
  \label{FictionistHomodyneObservable}
d^{\dag}d - c^{\dag}c = a^{\dag }b+b^{\dag }a \,,
\end{equation}
It is easy to verify that $\left\langle \psi _{n,\phi }\right\vert
a^{\dag }b+b^{\dag }a\left\vert \psi _{n,\phi }\right\rangle
\propto \sin \phi $. \ Thus, even the state $\rho _{ab}$ shows
interference, because every term in the incoherent sum is
proportional to $\sin \phi$.

So, the interference is explained by the fact that one has coherence between
different ways of distributing $n$ photons between a pair of modes, not by
the fact that one has coherence between different numbers of photons in a
single mode.

\textbf{Factist:} Well, I admit I can't see any mistake in what you've done,
but I'm still not convinced. Hasn't it been shown experimentally \cite{Pfl67}
that interference is obtained even between two \emph{independent} lasers?

\textbf{Fictionist:} Yes, but the interference that is observed
can still be explained without needing to invoke coherence. \ Let
me convince you of this using the simplest example of a pair of
Fock states. \ Suppose that initially the state is $\left\vert
n\right\rangle \left\vert n\right\rangle .$ \ The output port in
which the first photon is detected is completely random. \
However, after this detection, the state must be updated to
$(1/\sqrt{2})(\left\vert n-1\right\rangle \left\vert
n\right\rangle \pm \left\vert n\right\rangle \left\vert
n-1\right\rangle) $ with the relative phase being fixed by the
random outcome of the first detection. After many such detections,
the state evolves to the sort of state we have in the homodyne
experiment: a coherent superposition over different relative
photon numbers with a well-defined phase.\footnote{ The
 surprising result that two Fock states
yield an interference pattern in the joint distribution of a
multi-particle detection was first discovered in the context of
Bose-Einstein condensates by Javanainen and Yoo~\cite{Jav96}. The
issue was investigated in the optical context by M\o
lmer~\cite{Mol97}, and a simple analytical investigation can be
found in~\cite{Cab05}.} As argued
previously, such a state shows interference despite the fact that
the reduced density operators have no coherence.  Thus, one can
explain the interference of independent lasers without invoking
coherence.

\textbf{Factist}: Hmm. I'm sure there's \emph{some} example that
demonstrates the need for coherence. Otherwise, how could you explain the
fact that the predictions that were made on the basis of assuming coherent
states were never found to be in error? I just need to think some more about
it... What about the following case? [...]

[\textit{A long series of examples and fictionist explanations of
these examples follows.}\footnote{ See~\cite{San03} for the
fictionist response to many other standard interference
experiments.} \textit{Finally, the fictionist sees a pattern.}]

\begin{figure}
\begin{center}
\includegraphics[width=3.25in]{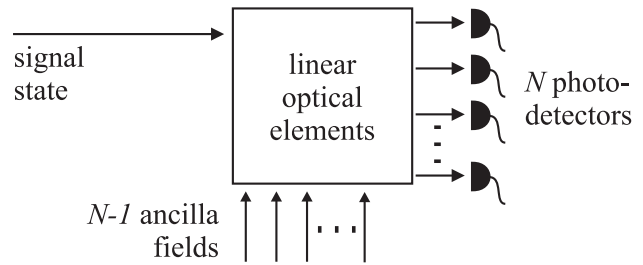}
\end{center}
\caption{A schematic for a general linear optical interferometer.
The signal mode and $N-1$ probe fields are injected into a general
linear optical interferometer, followed by photodetection on all
$N$ output modes.} \label{interferometer}
\end{figure}

\textbf{Fictionist: }You can stop looking for more examples,
because I have a general theorem that will deal with all linear
optical experiments.  The most general such experiment is a
$2N$-port interferometer, wherein one of the input ports
corresponds to the signal field (the one which we are trying to
identify as coherent or not), while the other $N-1$ correspond to
independent probe fields, and at each of the $N$ output ports is a
photodetector. The body of the interferometer may involve any
combination of linear optical elements [\textit{see
Fig.~\ref{interferometer}}].

Now consider the following. If the photodetectors were ideal, so
that together they constituted a measurement of the total number
of photons in all modes, and if the optical elements were
lossless, so that they conserved the total number of photons, and
if the probe fields were described by number states, so that the
total number of photons in the probe fields was known, then one
could immediately infer the number of photons in the signal mode
by taking the difference of the total number detected and the
total number in the probe fields. Such an idealized interferometer
would therefore constitute nothing more than a fancy measurement
of the number basis on the signal field.

In practice, photodetectors constitute an error-prone measurement of the
number basis \cite{Vogel}, photons may be lost to absorption somewhere in
the interferometer (which has the same effect as coupling into an output
mode upon which a measurement of the number basis is performed but the
outcome is unregistered), and each of the probe fields is, by M\o lmer's
argument, an incoherent mixture of number states. However, this simply means
that there is uncertainty in the number of photons in the probe and
uncertainty in the total number measured, and consequently that such an
interferometer is simply an error-prone measurement of the number in the
signal. But the statistics of such a measurement are still completely
insensitive to any off-diagonal elements of the density operator in the
number basis.

This result, by the way, explains why calculations wherein sources
are represented by coherent states have agreed so well with
experiments. The most general linear detection scheme (as described
above) is completely insensitive to the values of the off-diagonal
elements of the density matrix! One could assume \emph{any values
whatsoever} for these elements without affecting the result of such
calculations. The use of coherent states in the place of Poissonian
mixtures of number states will yield the correct predictions, and
may even simplify the calculation, but they are simply a convenient
fiction.

\textbf{Factist}: Fine. I grant that these sorts of experiments
don't settle the issue in my favor. But they don't decide it in
your favor either. All you've shown is that if you can't generate
coherence, then you can't detect it either. But if you
\emph{could} generate it, then you could also detect it.  I can
prove this to you using the simple homodyne example from before
where I imagine that the local oscillator comes from this
hypothetical coherent source.

Let the state of the local oscillator be a coherent state $\left\vert \beta
\right\rangle .$ \ A simple calculation shows that the difference in number
at the two output ports is
\begin{align}
  \text{Tr}_{ab}\Bigl[ e^{i\phi N}\rho e^{-i\phi N} \otimes & |\beta
  \rangle \langle \beta |
  \left( a^{\dag }b+b^{\dag }a\right)\Bigr]  \notag \\
  & = \text{Tr}_a \Bigl[ \rho (\beta^* e^{-i\phi }a+ \beta
  e^{i\phi}a^{\dag })\Bigr]\,.
\end{align}
Again, the only way one can obtain interference in this case is if
$\rho $ has coherence between different number eigenstates.
Therefore such a measurement is a test for the presence of coherence
in the signal. So this shows that if you \emph{could} generate
coherence, somehow, then you could use it to detect coherence.  So
it ultimately just comes back to the issue of whether you can
generate coherence,~...

\textbf{Fictionist: }But M\o lmer's argument...

\textbf{Factist: }... and I've now got a new idea for how to do
it. Basically, we just downconvert from an EM field that is of
sufficiently long wavelength.  Take a radio wave as an extreme
example.  Surely radio waves are in coherent states because the
way I generate them is by an oscillating current in an antenna
rather than by stimulated emission in atoms.  This current can be
treated classically.  It's just a charge moving up and down.
[\textit{Emphatically waves a fist up and down to illustrate.}] If
you look in Jackson~\cite{Jac98}, you'll find that a classical
oscillating current interacting with a quantum EM field generates
a coherent state.  Even microwaves can be generated by oscillating
currents, so all I need to do to get a coherent state at optical
frequencies is to downconvert from the microwave regime.

\textbf{Fictionist: \ }Even if you knew the time of the first peak
of a microwave to some error that is small relative to the period of
a microwave, this would be converted into an optical field with the
same peak position and the same absolute error on the peak position,
which is large compared to the period of an optical wave.  This
means that you would have complete ignorance of the optical phase,
and thus no coherence.

\textbf{Factist: }Well, someday it should be technologically
feasible to build antennae that have currents oscillating at
optical frequencies.  You have to admit that from that day onwards
we would have optical sources that were genuinely coherent.

\textbf{Fictionist:} Would I? Hmmm.  [\textit{Considers the question
for a while.}] It seems to me that even in the case of an
oscillating current, I can think back to how it was made and see
that one only ever gets entanglement, never coherence.

\textbf{Factist:} But then you would have to admit that microwaves
and radio waves are not in coherent states either!

\textbf{Fictionist:} Yes, now that I\ think about it, that's right.
So let me refine my thesis: coherence between number eigenstates
isn't just a fiction for fields at optical frequencies, it's a
fiction for fields of \emph{any} frequency.

\textbf{Factist: }Ok... [\textit{exasperated}] So how are you going
to obtain the classical limit? \ Everyone knows that classical EM\
fields correspond to large-amplitude coherent states, and you do
need to recover the classical limit for optical waves, microwaves,
radio waves, and electric currents under the right conditions.

\textbf{Fictionist:} I suppose that we will simply have to rethink
the notion of quantum-classical correspondence. Maybe there is some
alternative way to obtain classical EM fields.  I don't quite see it
yet, I admit, but I'm confident it will work out.

\textbf{Factist:} This makes no sense to me.  Why not assume
coherence?  Then you have no problems with the classical limit.

\textbf{Fictionist:} Look, even if you do assume that the
electrons in an antenna are in a coherent superposition of energy
eigenstates and therefore have a well defined phase, \emph{you
don't know the phase}.  When you wave your fist up and down
[\textit{repeats the motion}], it suggests that you could directly
know the phase of the electron motion, but even for radio waves
it's too high in frequency for you to ever know it.

\textbf{Factist:}  Let's not go back to that!  You already conceded
that there's nothing wrong with an ignorance interpretation of a
\emph{proper mixture}.  I don't claim that anyone necessarily knows
the phase, but simply that there is a well-defined phase.

\textbf{Fictionist:} Look, this whole conversation is starting to make my
head hurt. The point is that all of my calculations come out right without
the need to introduce coherences, so there's no reason to assume they exist.
They are just metaphysical baggage.

\textbf{Factist:} I've also got a head-ache, all of my calculations
also come out right, and I still think that you're wrong.

\section{A Resolution}

The debate we presented was ultimately about whether the intrinsic
properties of a system are best described by a coherent or an
incoherent quantum state. But the whole debate presumes that
quantum states only contain information about the \emph{intrinsic
properties} of a system.  We submit that this presumption is
mistaken; quantum states also contain information about the
\emph{extrinsic properties} of a system, that is, the
\emph{relation} of the system to other systems external to it, and
whether or not coherences are applicable depends on the external
system to which one is comparing.

The philosophical distinction between intrinsic and extrinsic
properties can be explained as follows \cite{Lew83}:
\begin{quote}
A sentence or statement or proposition that ascribes intrinsic
properties to something is entirely about that thing; whereas an
ascription of extrinsic properties to something is not entirely
about that thing, though it may well be about some larger whole
which includes that thing as part. A thing has its intrinsic
properties in virtue of the way that thing itself, and nothing
else, is. Not so for extrinsic properties, though a thing may well
have these in virtue of the way some larger whole is. The
intrinsic properties of something depend only on that thing;
whereas the extrinsic properties of something may depend, wholly
or partly, on something else.
\end{quote}

In the context of the argument of the dialogue, our claim is that
the phase of an optical mode is an extrinsic property of that mode,
only defined in relation to an external phase reference, and the
fact that there are many possible choices for this phase reference
is the source of the debate.

An analogy is useful here. In special relativity, if one is not
careful, an apparent contradiction arises. Suppose Alice and Bob
are observers with some non-zero relative velocity, and suppose we
denote the length of Alice's metre stick by $L_{1}$ and the length
of Bob's metre stick by $L_{2}.$ \ Now, a simple application of
length contraction would seem to suggest that Bob ought to
conclude that $L_{1}<L_{2},$ while Alice ought to conclude that
$L_{1}>L_{2}.$ But these statements are contradictory! Who is
right? Is it Alice's or Bob's metre stick that is longer? Surely,
one might argue, there must be some matter of fact about which is
longer. But the idea that either Alice or Bob must be right and
the other wrong is simply mistaken. The mistake was in assuming
that there is a single intrinsic property -- the length of Alice's
metre stick -- to which they are both referring. Really, there is
only the world-sheet of Alice's metre stick, and the world-sheet
of Bob's metre stick, and different time-slices of these sheets.
When Alice compares lengths she is comparing a particular pair of
time-slices of these sheets, while when Bob does so, he is
comparing a different pair of time-slices. Thus, if $L_1^{A}$ and
$L_2^{A}$ denote the lengths of the time-slices compared by Alice,
and $L_1^{B}$ and $L_2^{B}$ denote the lengths of the time-slices
compared by Bob, we have $L_1^{A}>L_2^{A}$ and $L_1^{B}<L_2^{B}$
and no contradiction. The appearance of a contradiction is
dispelled when one realizes that Alice and Bob were making claims
about different entities.

We propose the same sort of resolution to the dispute between the
factist and the fictionist. \ The factist's use of the coherent
state $\left\vert \alpha \right\rangle =\sum_{n=0}^{\infty
}\frac{e^{-|\alpha |^{2}/2}\alpha ^{n}}{\sqrt{n!}}\left\vert
n\right\rangle $\strut\ and the fictionist's use of the incoherent
state $\rho =\sum_{n=0}^{\infty }\frac{e^{-|\alpha |^{2}}|\alpha
|^{2n}}{n!}\left\vert n\right\rangle \left\langle n\right\vert $
seem to be at odds with one another because there is a presumption
that the two are describing \emph{the same degree of freedom,}
namely, the intrinsic properties of a single optical mode.

However, the dispute can be resolved if one grants that their
quantum states describe the relation between this optical mode and
an external phase reference, and if one recognizes that the
factist and the fictionist are implicitly making use of
\emph{different} external phase references.

Denoting the factist and fictionist's external phase references by
$R$ and $R^{\prime }$ respectively, we can make the point as
follows: whereas the factist's $\left\vert \alpha \right\rangle$
concerns \emph{the relation between} $S$ \emph{and} $R$, the
fictionist's $\rho$ concerns \emph{the relation between} $S$
\emph{and} $R^{\prime}$!  They are describing different entities
and so it is not a contradiction if their descriptions differ. In
the following, we shall attempt to defend this point of view in
more detail.

\section{Preliminaries about reference frames}

We begin with a few comments about reference frames (RFs). Because
the example of a Cartesian RF is arguably more intuitive than that
of a phase reference, we shall illustrate the central concepts
with this example.

When a quantum state of a spin-1/2 system is assumed to be spin-up
along the $\hat{z}$ direction, we are assuming the existence of a
Cartesian reference frame, with respect to which the $\hat{z}$
direction is defined. We need not assume that this Cartesian frame
is defined by Newton's absolute space, because we only ever
compare the orientations of physical objects to other physical
objects and never to any purported absolute space. \ Similarly, we
only ever compare the phases of optical modes to other oscillating
systems, and never to any purported absolute time standard, so we
have no need in practice of an absolute time standard.

Consequently, a reference frame can in practice always be taken to
be defined by some physical object. \ It follows that one ought to
be able to apply quantum theory to the reference frame itself if
one wishes. \ So in describing any given experimental situation,
one is forced to make a choice about whether the RF is treated as
\emph{external} or as \emph{internal}.  To be precise, to treat an
RF \emph{externally} is to treat it as a background resource to
which one's description of the system is referred. \ On the other
hand, to treat it \emph{internally} is to incorporate it into the
formalism and to assign it degrees of freedom like any other
physical system.

In quantum theory, treating an RF internally requires introducing
a Hilbert space for it.  Treating it externally usually implies
that it is being treated classically. However, for any theory, not
just quantum theory, one can introduce a distinction between
treating a reference system as part of the system under
investigation and treating it as part of the background. \ For
instance, in Newtonian mechanics, if we consider a ball bouncing
off a wall, we may treat the wall either as an external potential
or as a dynamical system that also obeys Newton's laws. We shall
be arguing that neither method of representation is preferred; it
is not as if one of these ways of treating a reference frame is
correct and the other incorrect. \ It is simply a conventional
choice of the physicist.

Another consequence of reference frames being defined by physical
objects is that there can be many distinct physical systems that
define reference frames for the same symmetry group. \ Thus, our
spin-1/2 system may be known to have been generated by
post-selecting the up outcome in a Stern-Gerlach experiment, so
that it may be said to be in the quantum state $\left\vert
+z\right\rangle $ relative to a $\hat{z}$-axis defined by the
Stern-Gerlach magnet. However, any other Stern-Gerlach magnet also
defines a $\hat{z}$-axis and if the two magnets are not aligned,
the spin-1/2 system will not be described by the state $\left\vert
+z\right\rangle $ relative to the second magnet. If the second
magnet is related to the first by a rotation $\Omega \in SO(3),$
then the quantum state of the spin-1/2 system relative to the
second magnet will be $R(\Omega )\left\vert +z\right\rangle $,
where $R(\Omega )$ is a unitary representation of $\Omega $. If we
consider a third magnet, for which the orientation to the first
magnet is completely unknown, then we must average over SO(3)
rotations with the uniform measure $d\Omega$ over SO(3), and the
quantum state of the spin-1/2 system relative to the third magnet
will be $\int d\Omega R(\Omega )\left\vert +z\right\rangle
\left\langle +z\right\vert R(\Omega)^{\dag }$~\cite{BRS03}.

It is worthwhile to introduce a distinction between an RF that has
some correlation with the system of interest, which we shall call
an \emph{implicated} RF, and one that is completely uncorrelated
with the system, which we shall call a \emph{nonimplicated} RF. In
our example of the spin-1/2 system, the first and second magnets
were implicated RFs while the third was a nonimplicated RF.

\begin{figure}
\begin{center}
\includegraphics[width=3.25in]{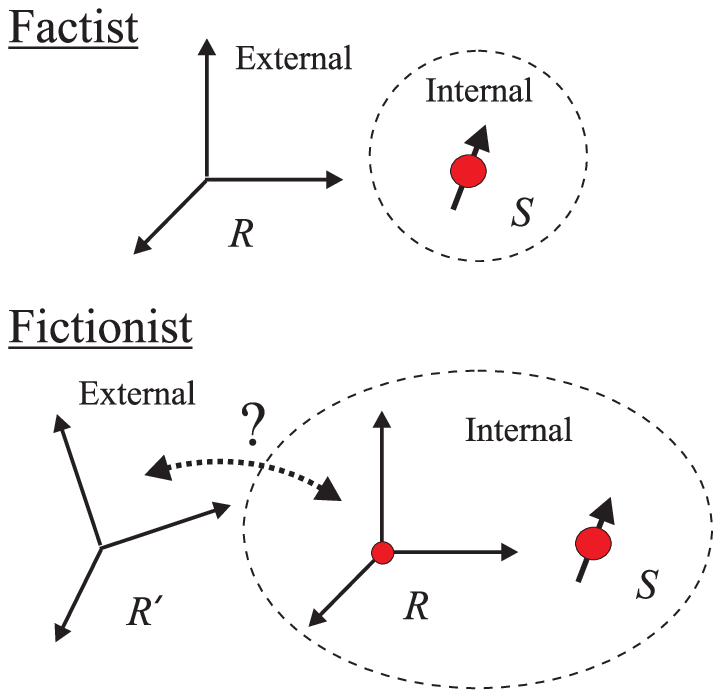}
\end{center}
\caption{Diagrammatic representation of the description of a system
and reference frame.  The factist treats the reference frame $R$ as
external, whereas the fictionist treats the reference frame $R$ as
internal and uncorrelated to any external reference frame $R'$.}
\label{RFs}
\end{figure}

We now repeat the main point of our resolution to the controversy
for this case.  The factist's description of a system $S$ is of
its relation with an external Rf $R$ that is implicated, whereas
the fictionist, who insists on internalizing $R$, describes $S$ in
terms of its relation with an external RF $R'$ that is
nonimplicated.  Because they are describing different entities,
there is no contradiction if the quantum states they use differ.
Fig.~\ref{RFs} illustrates the two paradigms of descriptions.

\section{Reconsidering the dialogue according to our resolution}

In the dialogue, the factist and fictionist both made an
assumption which we believe to be mistaken, namely, that their
quantum states describe only intrinsic properties of a system. If
this assumption is relaxed, then the differences in their
convictions can be accounted for reasonably well by supposing that
they are simply disagreeing about the RF with respect to which the
extrinsic properties of the system are being defined. Essentially,
the factist is inclined towards treating the RF to which the
system is correlated as an external RF, while the fictionist is
inclined towards treating this RF internally, and leaving only a
nonimplicated RF as external.

\subsection{Making sense of the states in the dialogue}

We shall consider the homodyne experiment described in the dialogue
(but we start by focusing on the states rather than the
measurements). Here, the system $S$ is the signal mode, and the
reference frame $R$ is the local oscillator. When this is first
discussed in the dialogue, the factist describes $S$ by the coherent
state $|\alpha \rangle$, and treats $R$ as a classical field.
Because $R$ has no representation in the quantum formalism, the
factist is treating his phase reference externally. The fictionist,
on the other hand, insists on treating $R$ within the quantum
formalism. Consequently, $R$ is an \emph{internal} phase reference
for her. But the fictionist still implicitly makes use of an
external phase reference, which we shall denote by $R'$; it is just
that the relation between $R'$ and $R$ is assumed to be completely
unknown.

(One might argue that the fictionist need not have an external
phase reference at all. However, our fictionist agrees with the
factist on the mathematical structure of the Hilbert space,
including the formal possibility of coherence, i.e., she agrees
that phase is ``in principle'' an observable on this system.  This
suggests that she does possess an external phase reference but it
is simply uncorrelated with $S$ or $R$.)

We now demonstrate that the incoherent state that the fictionist
assigned to $S,$ namely, the Poissonian mixture of number states
of Eq.~(\ref{incoherentstate}) is precisely how one would describe
the relation between $S$ and external $R^{\prime }$ given that one
describes the relation between $S$ and external $R$ by the
coherent state $\left\vert \alpha \right\rangle $ of
Eq.~(\ref{coherentstate}) and given that the relation between $R$
and $R^{\prime }$ is completely unknown. If $R^{\prime }$ was
related to $R$ by the phase $\phi \in [0,2\pi)$, and if the
relation between $S$ and $R$ was described by the quantum state
$\sigma$, the relation between $S$ and $R^{\prime }$ would be
described by $U(\theta )\sigma U(\theta)^\dag$, where $U(\theta
)=e^{i\theta N}$ acts unitarily on the Fock space
$\mathcal{H}_{S}$ of a single mode, and where $N$ is the number
operator on $\mathcal{H}_S$~\cite{Enk05}. However, if one has no
knowledge of the phase $\theta$ then one must average over all
$\theta \in [0,2\pi)$~\cite{BRS03}, implying that the relation
between $S$ and $R^{\prime}$ is described by
\begin{equation}
  \int_0^{2\pi} \frac{d\theta}{2\pi}\, U(\theta )\sigma U(\theta )^{\dag }\,.
\end{equation}

Thus, if we assume that the relation between $S$ and external $R$
is described by the factist's quantum state, $\left\vert \alpha
\right\rangle \left\langle \alpha \right\vert ,$ the relation
between $S$ and external $R^{\prime }$ ought to be described by
the quantum state
\begin{align}
  \int_0^{2\pi} \frac{d\theta}{2\pi}\, U(\theta )\left\vert \alpha \right\rangle
  \left\langle \alpha \right\vert U(\theta )^{\dag }  &= \int_0^{2\pi}
  \frac{d\theta}{2\pi}\,
  \left\vert \alpha e^{i\theta }\right\rangle \left\langle \alpha
  e^{-i\theta }\right\vert
  \nonumber \\
  &=\sum_{n=0}^{\infty }\frac{e^{-|\alpha |^{2}}|\alpha
  |^{2n}}{n!}\left\vert n\right\rangle \left\langle n\right\vert \,,
\end{align}
where the final equality is the familiar ambiguity of mixtures
discussed in the dialogue.  But this is precisely the quantum
state assigned by the fictionist.

At one point in the dialogue, the factist argues that the reduced
density operator on $S$ of the fictionist's quantum state for
$R+S$ has a preferred convex decomposition into coherent states,
and in this sense is consistent with the factist's quantum state
for $S$. The fictionist rightly points out that the factist cannot
simultaneously agree that $R$ and $S$ are entangled while
maintaining that $S$ is in a pure state.  By the lights of our
account, the factist description \emph{is} consistent with the
fictionist's, but for a very different reason. What the factist
describes as ``the state of $S$'' is really the quantum state
describing the relation between $S$ and $R$, and what the
fictionist describes as ``the state of $S$'' is really the quantum
state describing the relation between $S$ and $R'$.  Therefore, if
we want the fictionist's description of the relation between $S$
and $R$, we don't want what she calls "the reduced density
operator on $S$" nor any element of a convex decomposition
thereof. What we really want is the quantum state on a Hilbert
space that somehow encodes the \emph{relation} between the Hilbert
space she associates with $S$ and the Hilbert space she associates
with $R$. As it turns out, it is the noiseless
subsystems~\cite{KLV00} with respect to phase rotations that serve
this purpose.  The states within these noiseless subsystems are
coherent states in the limit of a large phase reference. See
Ref.~\cite{BRS04a} for more details.

In the dialogue, the argument over coherence was often repeated
``one level up'', for instance, at the level of the gain medium of
the laser rather than the field. Arguments at these higher levels
can be understood in the same way as the arguments at the lower
level. To see this, it is useful to redescribe what occurred in the
dialogue in the light of our resolution.

After the initial argument over coherence of the system $S$ (the
field), the factist mistakenly buys into the fictionist's argument
and agrees to internalize $R$ (the gain medium), but he still
insists on describing the pair relative to some external $R^{*}$
that is correlated to $R$ (for instance, the electrons of the
pumping mechanism, considered as an external RF).  The fictionist
does not agree to allow $R^{*}$ as an external RF and so promptly
internalizes it, and continues to describe everything relative to an
uncorrelated RF $R'$, which is left implicit in the discussion.
Again, both descriptions are equally valid and differ only insofar
as they relate the system under investigation to different external
RFs.

As a specific example, we consider the disagreement that arises in
the case where the RF $R$ is the local oscillator in the homodyne
experiment. When the factist internalizes the local oscillator
(which occurs when he makes the case that one can detect coherence
given a coherent source), he describes it relative to an
\emph{implicated} external RF $R^*$.  In the dialogue, the factist
assumes an $R^*$ which is aligned perfectly with $R$. Thus, the
factist describes the relation between $S$ and $R^*$ with
precisely the same quantum state as he used to describe the
relation between $S$ and $R$, namely, $|\alpha\rangle$.  The
factist describes the relation between $R$ and $R^*$ by a coherent
state $|\beta\rangle$.

Now consider how to redescribe $R+S$ relative to a nonimplicated
reference frame $R'$.  Let the system be realized by mode $a$ and
the reference $R$ by mode $b$.  Given that the relation between $R'$
and $R^*$ is completely unknown, we must average the factist's
quantum state over all phase rotations on $R+S$.  Thus, the quantum
state on modes $a$ and $b$ relative to $R'$ is
\begin{equation}
\label{rho_ab1}
    \rho_{ab} = \int_0^{2\pi} \frac{d\theta}{2\pi}\, V(\theta) |\alpha \rangle
    \langle \alpha| \otimes |\beta \rangle \langle \beta|
    V^{\dag} (\theta)\,,
\end{equation}
where we have defined the unitary operator $V(\phi) = \exp(i\phi
N_b)\exp(i\phi N_a)$, with $N_{a,b}$ the number operators for modes
$a,b$.  This state is equal to
\begin{equation}
  \label{rho_ab2}
  \rho_{ab} = \sum_{n}p_{n}| \psi_{n,\phi}\rangle
  \langle \psi _{n,\phi}|\,,
\end{equation}
where $p_n$ is a Poissonian distribution over $n$ and
$\psi_{n,\phi}$ is defined in Eq.~(\ref{psi_n_phi}), where the
parameter $T$ which appears therein is related to $\alpha$ and
$\beta$ by $|\alpha|^2/|\beta|^2= (1-T^2)/T^2$.  This can be
easily verified using the techniques of Ref.~\cite{San03}. But
this is precisely the state, Eq.~(\ref{rho_ab}), that is adopted
by the fictionist to describe $R+S$.

Whenever a reference frame $R$ is internalized, one's description
becomes relative to a new \emph{external} reference frame.  If,
however, in the experiments of interest, the system $S$ is only ever
compared to $R$, and no comparison of either is ever made to the new
external reference frame, then for the predictions of the outcomes
of such experiments, it makes no difference whether the new
reference frame is implicated or not, and thus it makes no
difference what the distribution over the global phase of $R+S$ is.
For such experiments, the factist's and the fictionist's
descriptions yield completely equivalent predictions.

Note, in addition, that whenever one internalizes a reference
frame, one must choose a physical description (i.e., a quantum
state) to represent what was previously described classically.
Thus, one must deal with the myriad of issues associated with the
quantization of a classical system.  In particular, if it is
demanded that the new description (with an internal RF) gives
identical predictions as the previous description (with an
external RF), the quantum state chosen must satisfy some
conditions of a ``classical limit''.  For example, when the
factist is convinced by the fictionist that he should treat his
local oscillator internally, he chooses to represent the local
oscillator quantum mechanically as a coherent state
$|\beta\rangle$ (relative to his new external phase reference
$R^*$).  It would seem necessary, then, that he takes the $|\beta|
\rightarrow \infty$ limit in order to have complete agreement with
his previous description.  We will return to this issue in the
next subsection, when we consider measurements.

\subsection{Making sense of the measurements in the dialogue}

To further our case, we show that the observable $ba^{\dag} +
b^{\dag}a$ of Eq.~(\ref{FictionistHomodyneObservable}) that the
fictionist uses to describe the homodyne measurement is precisely
the observable describing a measurement of the quadrature of the
signal \emph{relative to} the local oscillator when the latter is
treated internally, all from the perspective of a nonimplicated
external RF.  We establish this equivalence in two steps.

First, we note that the observable $ba^{\dag }+b^{\dag }a$ on two
modes ($a$ and $b$) is the same with respect to \emph{any}
external phase reference, because it is invariant under
``passive'' phase shifts (that is, the unitary operator $V(\phi) =
\exp(i\phi N_b)\exp(i\phi N_a)$ as defined in the previous
section).  We say, then, that the observable $ba^{\dag} +
b^{\dag}a$ only yields information about the relative degrees of
freedom of $S+R$, because it is independent of any external RF.
Moreover, any party with a \emph{nonimplicated} RF claiming to
measure only relative degrees of freedom we argue must use
observables with this invariance property.\footnote{The problem of
determining the \emph{optimal} measurements for inferences about
relative degrees of freedom is considered in
Refs.~\cite{BRS04b,BLS05,Gis05}}

Second, we show that the fictionist reproduces the factist's
predictions, which implies that the fictionist is also
implementing a measurement of the quadrature of $S$ relative to
$R$.  Recalling that the factist describes the measurement by
$\beta^* a + \beta a^\dag$ and describes the relation between $S$
and the external $R$ (after the phase shifter) by the coherent
state $|\alpha e^{-i\phi}\rangle$, it follows that he predicts a
mean quadrature
\begin{equation}
  \langle \alpha e^{-i\phi} | (\beta^* a + \beta a^\dag) |
  \alpha e^{-i\phi} \rangle = \beta^* e^{-i\phi}\alpha + \beta
  e^{i\phi}\alpha^* \, .
\end{equation}

The fictionist, on the other hand, describes $R+S$ relative to a
nonimplicated external RF $R'$ using $\rho_{ab}$ of
Eq.~(\ref{rho_ab}), which can be written as a mixture of products
of coherent states, as in Eq.~(\ref{rho_ab1}). After the phase
shifter, we have
\begin{equation}
    \rho_{ab}(\phi) = \int_0^{2\pi} \frac{d\theta}{2\pi}\,
    V(\theta) |\alpha e^{-i\phi} \rangle
    \langle \alpha e^{-i\phi}| \otimes |\beta \rangle \langle \beta|
    V^{\dag} (\theta)\,,
\end{equation}
Thus, the mean value of $a^{\dag}b+ab^{\dag}$ is
\begin{equation}
  \text{Tr}\bigl[ \rho_{ab}(\phi) (a^{\dag }b+ab^{\dag })\bigr]= \beta^*
  e^{-i\phi}\alpha + \beta e^{i\phi}\alpha^* \, .
\end{equation}
This coincides with the factist value for the mean quadrature, so
it is appropriate to say that the fictionist's measurement is
indeed of the relative quadrature of $S$ to $R$.

Note that achieving this agreement did not require the amplitude
of the coherent state $|\beta\rangle$ to be large. However, if
\emph{all} measurements (such as higher-order correlations of the
photocurrents) are to be equivalent, one must take the limit
$|\beta|\rightarrow \infty$. This would be required, for example,
if one performed the above analysis using the formalism of
generalized measurements (POVMs~\cite{Nie00}) and demanded that
the Born rule gave equivalent results in both descriptions. If
$|\beta|$ was finite then there would be correction
terms~\cite{Tyc04}.

In the dialogue, the argument about whether it is possible to
detect coherence ends with the factist concluding that  ``if one
could generate coherence then one could detect it''. What the
factist really establishes with his argument is simply that if an
internal RF $R$ (the local oscillator) is correlated with an
external RF $R^*$ (so that the quantum state of the local
oscillator has coherence), then by measuring the relation between
$S$ and $R$ (through the homodyne detection), one obtains
information about the relation between $S$ and $R^*$, in
particular, whether they are correlated (and thus whether the
quantum state of the signal was coherent or not).

Finally, it should be noted that if one is measuring only
relations among the parts of the system, then the relation between
the system and the external RF will be of no significance.
Mathematically, such measurements are associated with incoherent
POVMs, i.e., POVMs that are invariant under collective phase
rotations, so that the statistics are insensitive to the phase
distribution of the state.  This has been emphasized by Nemoto and
Braunstein~\cite{Nem02}. Notwithstanding this fact, it is always
possible to make a measurement of the relation between the system
and an external RF, and in this case the POVM is non-invariant
under phase rotations and the phase distribution of the state has
empirical significance. As an example, in the factist's original
description of homodyne detection, where the quantum state
describes the relation between the system and an external local
oscillator, the use of a coherent state was necessary to obtain
the correct predictions.

\subsection{Making sense of the transformations in the dialogue}

In the dialogue, the factist and the fictionist both described the
phase shifter as a transformation of the intrinsic properties of
the system.  However, it is only the relation between the signal
and local oscillator that is affected. This view is corroborated
by the fact that the outcome of the measurement would be precisely
the same if the phase shifter was instead placed in the path of
the local oscillator mode.  In other words, although active and
passive transformations may be distinguished in the formalism,
there is no physical distinction between them.

\subsection{Relative localization}

We can also understand the results on the interference of two
lasers within our perspective.  Consider one of the lasers to be
the signal and the other to be the phase reference. To say that
they are independent is to say that the phase reference is not
correlated with the signal at the outset, i.e. nonimplicated in
our terminology.  The fact that interference can be achieved after
sufficiently many photodetections demonstrates that the two become
correlated over time.  But this is precisely what one expects
given that the homodyne detection implements a measurement of the
relative phase of the two; the quantum state is updated to one
that reflects the particular relative phase measured in the
experiment. Note that the same conclusion could be reached if one
of the lasers was treated externally (i.e. classically). The
relation between the two modes would be described by a mixture of
coherent states, and the phase distribution would evolve, as one
accumulated data, from uniform to highly peaked about some random
value.  In other words, the external phase reference would evolve
from nonimplicated to implicated. See~\cite{Cab05} for further
details.

\subsection{Microwaves, radio waves, and humanly-perceptible
oscillations}

At the end of the dialogue, the fictionist argues against the
factist's suggestion that upconversion from coherent sources at
low frequencies could yield coherence at optical frequencies; this
argument is originally due to Wiseman~\cite{Wis04}.  Thereafter,
the factist appeals to a hypothetical optical frequency antenna,
which leads the fictionist to deny that even radio waves or
microwaves could be coherent. Essentially, the factist was
appealing to the notion that a reference frame consisting of
oscillating electrons is somehow more worthy to be left as an
external RF than is a RF consisting of a high intensity EM field.
The fictionist eventually decides to stick to his program of
constant internalization of any RF and denies that even this sort
of RF can be left external.\footnote{In some of the literature on
this debate~\cite{Rud01a,vEF02a,Spe03} there was an impression
that an optical frequency antenna was somehow a more legitimate
clock than a laser. The notion that the fictionist arguments apply
equally well to electronic clocks as to optical clocks was first
made by Wiseman~\cite{Wis04}.}  There is only one holdout by the
end of the dialogue, which is an observer's own sense of time. The
fictionist essentially argues that the only thing that she is
willing to take as an external RF would be a human's own internal
clock.\footnote{The idea that this might be the only clock that a
die-hard fictionist would allow to be external (and that the
reason she would allow it is the persuasiveness of waving one's
hand up and down!) is also due to Wiseman~\cite{Wis04}.}  However,
the choice of whether to treat \emph{any clock} internally or
externally is a conventional one; the clock provided by an
observer's sense of time has no greater claim to being an external
RF than any other.

\subsection{Applying the arguments in the dialogue to the whole
universe}

The debate in the dialogue could have been continued to the point
where the factist and fictionist were arguing about the existence
of coherence in the initial state of the universe. Similarly, for
our proposed resolution, one could ask what occurs when the entire
universe is the quantum system of interest, such that there is no
physical system left over to act as an external RF.  To ask either
of these questions is to presume that it makes sense to apply
quantum theory to the universe as a whole. It appears to us highly
likely, however, that quantum theory, as it is currently
formulated, applies only to subsystems of the universe.  This is
not to say that we reject the idea of theories that apply to the
universe as a whole; we only reject the idea that quantum theory
as it stands is such a theory.

\section{The controversy in other contexts}

We now consider how our arguments generalize to other contexts in
which the quantum coherence controversy arises.  We begin by
noting the connection to superselection rules.

To assert that there is a selection rule for some quantity is to
assert that this quantity is conserved. On the other hand, to
assert that there is a \emph{superselection rule} for a quantity
is to assert the impossibility of preparing coherent
superpositions of nondegenerate eigenstates of the associated
Hermitian operator (more generally, the impossibility of preparing
density operators with elements connecting nondegenerate
eigenstates) \cite{WWW52}.

The position of the fictionist can be restated as a belief in the
existence of a superselection rule for total photon number. We
have argued that whether to use a factist or fictionist
description is a conventional choice that depends on one's choice
of RF.  This implies that whether to adopt a superselection rule
for photon number is also a conventional choice that depends on
one's choice of RF. More specifically, our claim is that a
superselection rule for photon number is applicable if and only if
one's external phase reference is nonimplicated.

It is tempting to think that this idea -- that it is a
conventional choice of the theorist that determines whether a
superselection rule holds or not -- can only be maintained in some
contexts but not others. For instance, one might think that
coherence between charge eigenstates is different in kind from
coherence between photon number eigenstates.

However, we do not see any significant difference between photon
number and other conserved quantities. Admittedly, it may be more
difficult to construct good reference frames for some degrees of
freedom, but there is nothing in principle preventing their
construction. For instance, to lift the superselection rule
associated with charge, one must simply have a large reference
system with respect to which one can coherently exchange charge,
as argued by Aharonov and Susskind \cite{Aha67}.  As another
example, the experimental realization of Bose-Einstein
condensation in alkali atoms provided a reference frame for the
phase that is conjugate to atom number. When treating this RF
externally, interference experiments with condensates must be
interpreted in terms of states that are coherent superpositions of
eigenstates of atom-number.  We see no obstacle in principle to
lifting more general sorts of superselection rules as well.

Similarly, one might think that there are conserved quantities for
which a superselection rule is never applicable. For instance, the
case of linear momentum may appear to be different in kind from
that of photon number, because a superselection rule for linear
momentum would seem to imply that objects could not be localized
in space, and \emph{this}, one might think, would be contrary to
what is observed. However, all that is ever observed is the
localization of systems relative to other systems. If we treat
these reference systems internally, and refer our systems to a
nonimplicated external RF, then we have a paradigm of description
wherein there is no coherence between eigenstates of total linear
momentum, and thus a superselection rule for the latter, while
relative localization is still achieved. Such cases are not
different in principle from any other conserved quantity. What
sets them apart in practice is the ubiquitous nature of the
associated reference systems, such as those for spatial location.
The more ubiquitous a RF, the more theorists seem inclined to
treat it externally, but an internalized treatment is just as
valid.

Several recent papers have considered the problem of internalizing
reference frames. Specifically, it has been shown how to
internalize a Cartesian reference frame for spin
systems~\cite{Pou05}, a clock for spin systems~\cite{Pag83,Pou05},
and a clock for oscillators~\cite{PouMil05}.  A particular
strategy for internalization of general reference frames can be
found in Ref.~\cite{KMP04}.

\section{Neither party wins the debate}

In the end, which of our two protagonists can claim to have won
the debate?  Ultimately, treating one view as superior is only
justified if the alternative can be shown to lead to factual
errors, or to descriptions of such complexity that -- as for
Simplicio and Segredo in Galileo's Dialogue -- Occam's razor can
be brought to bear.

We have argued that if reference frames treated internally are
given ``classical limit'' descriptions compatible with an
externalized treatment, then the factist and the fictionist give
identical predictions for all experiments.  Thus, neither party's
state assignments can be held up as more empirically accurate than
the other's.  (In as much as this debate captures many aspects of
the controversy about continuous variable quantum
teleportation~\cite{Rud01a,vEF02a,Rud01b,vEF02b,Nem02,Wis03,Wis04,Smo04},
we believe that neither position in this controversy, factist or
fictionist, is more correct than the other.  See
also~\cite{BDSW04} for further insight into the issue of the
entanglement resource in continuous variable quantum
teleportation.)

Can we use Occam's razor to favor one description over the other?
The factist's description makes use of a smaller Hilbert space,
and thus in many situations may be more efficient, especially when
numerical calculations are required.

However, as mentioned, a limitation of our analysis is that we can
assert the empirical equivalence of the two viewpoints only in the
case where the physical systems comprising the reference frame
have consistent physical descriptions from both the factist and
fictionist points of view. In particular, this has necessitated
the use of large amplitude coherent states in describing the local
oscillators of the homodyne detection; these states are, in the
large amplitude limit, ``perfect'' reference frames.

However, physical reference frames are never perfect -- they are
finite, suffer back action and drift over time.  It may therefore
appear ``more correct'' to treat such RFs quantum mechanically,
and thus view the fictionist's description as superior to that of
the factist.

Nonetheless, it seems to us quite plausible that the imperfection
of physical reference frames may be taken into account within the
externalized mode of description of the factist by making use of
mixed rather than pure states, generalized measurements rather
than projective ones, and quantum operations rather than
unitaries.\footnote{It may also be necessary to use a Hilbert
space with superselection sectors.} For instance, Tyc and
Sanders~\cite{Tyc04} have determined the generalized measurement
for homodyne detection using a finite local oscillator.  Note,
however, that such descriptions, if they exist, are likely to be
justified theoretically by first considering a fully
quantum-mechanical model of the reference frame.

Thus, one of the positive outcomes of this debate and the
resolution presented herein is that we have identified a set of
interesting problems for future research: first, to determine the
effects of various imperfections in one's RF within the fully
quantum-mechanical treatment of the fictionist paradigm, and
second, to determine how these effects can be modelled within the
more economical paradigm of the factist by generalized quantum
operations. These problems are significant insofar as imperfect
reference frames may pose a challenge to achieving the fine
control that is required for the successful implementation of
quantum information processing protocols.

\begin{acknowledgments}
  The authors gratefully acknowledge Barry Sanders,
  John Sipe, and Howard Wiseman for many long discussions on these matters,
  and Sam Braunstein, Florian Girelli, Netanel Lindner, David
  Poulin, Lana Sheridan, Danny Terno and Howard Wiseman for their
  thoughtful comments on a draft
  of this paper. S.D.B. is supported by the Australian Research Council, and
  T.R.\ is supported by the UK Engineering and Physical Sciences
  Research Council.
\end{acknowledgments}

\appendix

\section{A comparison with alternative responses}
\label{app:comparison}

Wiseman~\cite{Wis04} was the first to have the key insight that
electronic and optical clocks are equally valid as phase
references. Nonetheless, he still insisted on internalizing his
clocks, and, unwilling to acknowledge any further external phase
reference, he was pushed to justify the use of a coherent state
for his clock by claiming that ``...if the laser itself is the
clock, then by definition it is coherent with respect to itself,
so that it should be described by a pure [coherent state] of zero
phase''. By the light of our analysis, this is a mistake.  No
quantity can be defined relative to itself.  In particular,
coherences describe relations -- and as such are only defined
between distinct systems.

The following are valid treatments of the clock: (1) the clock is
treated as an external RF, in which case it is not described by a
quantum state and therefore the issue of whether this state is
coherent or not does not arise; (2) the clock is treated as an
internal RF and the external RF with respect to which it is
described is implicated, in which case one can justify the use of a
coherent state for the internalized clock; (3) the clock is treated
as an internal RF but the external RF with respect to which it is
described is \emph{nonimplicated}, in which case one must use an
incoherent state to describe the internalized clock. When coherence
appears in the state of an internal RF it is with respect to an
external implicated RF, not with respect to itself.

Another response to the controversy, advocated by van Enk and
Fuchs~\cite{Fuc02}, is
that coherent
states play a privileged role in the description of the
propagating laser output state by making use of the quantum
di~Finetti theorem. Their arguments are compatible with the
fictionist's description, as the state they ascribe the laser has
no global coherence (when all modes are included).  They argue
that complete measurements on some of the propagating modes
relative to some external classical RF (the example they use is a
microwave field) will result in the remaining modes of the system
becoming correlated in phase to this external RF.  In our
language, their argument is that the fictionist can perform
complete measurements on part of the system relative to her
nonimplicated RF, correlating the systems and thus implicating the
external RF. Thus, their argument is essentially that the
fictionist could, in principle, perform measurements to implicate
her external RF and, in so doing, move to a factist description.

\section{The epistemic view of quantum states}

A final caveat is in order. When we say ``coherent quantum states
concern extrinsic properties'', we do not mean to suggest that the
quantum states of some relational degree of freedom are in
one-to-one correspondence with the different possible values that
this degree of freedom may take, that is, the different possible
physical states for that relational degree of freedom. We oppose
this view, and believe that it is far more likely that the quantum
states of a degree of freedom are in one-to-one correspondence
with the different possible \emph{states of knowledge} that one
can have about the value of that degree of freedom, even though a
satisfactory interpretation of the quantum formalism along these
lines has yet to be provided. (See Refs.~\cite{Fuc02,Spe04} for
discussions of this research program.) However, the argument we
presented here does not really depend on which of these two views
-- the ontic or the epistemic view -- one takes towards quantum
states. Whether quantum states describe states of reality or
states of knowledge about reality, we argue that they do not
simply concern intrinsic properties but extrinsic properties as
well.

\end{document}